\documentclass[aps,pre,twocolumn,showpacs]{revtex4} 
\usepackage{graphicx}
\usepackage{color}
\usepackage{hyperref}
\usepackage[justification=centering]{caption}
\usepackage[sans]{dsfont}
\usepackage{subfig}

\usepackage{hyperref}
\usepackage{dsfont}
\usepackage{babel}
\usepackage{amsmath}
\usepackage{amssymb}
\usepackage{lipsum}
\usepackage{amsmath}
\usepackage{amssymb}
\usepackage{lipsum}
\usepackage[sans]{dsfont}
\usepackage{multirow}
\usepackage{afterpage}
\usepackage{amsfonts}
\usepackage{mathrsfs}
\usepackage{natbib}
\usepackage{bm}%
\usepackage{babel}
\usepackage{hyperref}
\usepackage{graphicx}
\usepackage{color}
\usepackage{amsthm}
\def\fnum@figure{\figurename\thefigure}
\renewcommand{\figurename}{Fig.}

\begin{document}
\title{Ergotropy and Work Extraction in Quantum Heat Engines via Quantum Channels}
\author{Indrajith.V.S}
\email{indraphysics08@gmail.com}
\author{Disha Verma $^{\dagger}$}
\affiliation{ $^{*}$ Qdit Labs Pvt. Ltd, Bangalore - 560092, Karnataka India\\
 $^{\dagger}$ National Institute of Technology, Tiruchirapalli, Tamil Nadu, India.}
\bigskip

\begin{abstract}
    This paper explores quantum heat engines based on qubit and qutrit working media interacting with thermal environments through generalized amplitude damping (GAD) channels. We investigate how quantum channels can be employed to model heat absorption, dissipation, and work extraction in open quantum thermal machines, and derive the conditions required for positive work extraction. The effects of quantum correlations, emission probability, population redistribution, and system--environment interactions on the thermodynamic performance of the engine are systematically analyzed across different operational regimes. In addition, we examine the ergotropy of qubit and qutrit systems under dissipative dynamics to understand how environmental effects influence the maximum extractable work. Our results demonstrate that multilevel quantum systems exhibit enhanced work extraction capability and improved robustness against decoherence compared to two-level systems, providing further insight into the role of dissipative dynamics and quantum resources in realistic quantum thermodynamic devices.
\end{abstract}
\maketitle
\section{Introduction}
Quantum thermodynamics provides a fundamental framework for understanding energy, work, and information processing at the quantum scale, where thermal fluctuations and quantum effects coexist and strongly interact. It extends classical thermodynamic principles by incorporating uniquely quantum features such as coherence, correlations, and measurement back-action, thereby enabling a consistent description of nonequilibrium processes in microscopic systems. As a unifying discipline, quantum thermodynamics bridges fundamental physics and emerging quantum technologies, offering critical insights into the ultimate limits of efficiency, control, and performance of nanoscale devices.

Within this framework, several quantum-enabled technologies have been proposed and developed, including quantum thermal machines, quantum batteries, and quantum thermodynamic sensors \citep{qhe1,qhe2,qhe3,QHE_measureemnt, battery_1,battery_2, battery_3, battery_4, battery_5, campaioli2019quantum}. Among these, quantum heat engines and quantum batteries have attracted particular attention in recent years due to their potential applications in energy storage, energy conversion, and quantum information processing \citep{My_batetry_2,My_battery_1, My_QHE, kosloff2014quantum, goold2016role, quan2007quantum, allahverdyan2004maximal, vznidarivc2010thermalization, rossnagel2013nano}.

Quantum heat engines (QHEs) represent a compelling extension of classical heat engines into the quantum regime, where energy exchange and work extraction are governed by the laws of quantum mechanics \citep{qhe1}. A typical QHE consists of a microscopic working medium—such as a two-level system, harmonic oscillator, or many-body quantum system—that undergoes a thermodynamic cycle while interacting with thermal reservoirs. Unlike classical engines, the performance of QHEs is influenced not only by temperature gradients but also by genuinely quantum features such as coherence, quantum correlations, and finite-time effects. Consequently, QHEs serve as powerful platforms for probing the fundamental laws of thermodynamics at the quantum level and for assessing the role of quantum resources in energy conversion and information processing.

A broad class of QHE models has been proposed that exploit distinct quantum mechanical effects. For example, coherent thermal reservoirs can enhance energy transfer through the presence of quantum coherence, leading to improved engine performance \cite{coh_bath1, coh_bath2, coh_bath3, coh_bath5}. Squeezed thermal baths, which are inherently non-equilibrium quantum states, enable reduced entropy production and enhanced work extraction \cite{sq_bath1, sq_bath2, sq_bath3}. The incorporation of non-Markovian dynamics introduces environmental memory effects that help preserve coherence and mitigate dissipative losses \cite{non_mar_bath1}. Furthermore, quantum-correlated reservoirs, where entanglement exists between thermal baths, allow for energy extraction beyond classical thermodynamic limits  \cite{corr_bath1, Exceptional_points}.

Beyond these mechanisms, recent studies have investigated the roles of quantum measurements, feedback control, and coherence recycling in optimizing engine efficiency. Additional proposals explore many-body working substances, strong system–bath coupling regimes, and the exploitation of quantum phase transitions to enhance power output and efficiency, underscoring the rich landscape of quantum-enhanced thermal machines.

In parallel, significant efforts have been directed toward developing QHE models based on realistic and experimentally accessible physical systems, with the goal of bridging the gap between idealized theoretical constructs and practical implementations \citep{cooper_pair_tunnelling}. Representative platforms include deformed quantum fields \citep{QHE_deformed_1, QHE_deformed_2}, quantum spin chains \citep{QHE_Spin_Chain_1, QHE_Spin_Chain_2, unit_efficieny}, optomechanical systems \citep{Opto_mechanical}, and superconducting qubits \citep{QHE_Superconduct}.

Complementing these theoretical advances, experimental realizations of quantum heat engines have been successfully demonstrated across diverse platforms. Notable examples include implementations using single trapped ions \citep{QHE_Trpped_ion_1}, ensembles of nitrogen-vacancy (NV) centers in diamond \citep{QHE_NV}, and nuclear magnetic resonance systems \citep{QHE_Trpped_ion_1, QHE_Trpped_ion_2}. These experimental achievements provide valuable insights into quantum thermodynamic processes and validate key theoretical predictions.

More recently, measurement-based quantum heat engines have emerged as a promising paradigm, wherein quantum measurements themselves play an active role in energy exchange and work extraction \citep{Measurement_QHE_1, Measurement_QHE_2, Measurement_QHE_3, Measurement_QHE_4}. In particular, weak measurements enable partial information extraction with minimal disturbance, allowing the system to retain coherence while facilitating controlled energy flow \citep{QHE_weak}. Such engines highlight the fundamental thermodynamic role of information and measurement back-action, opening new avenues for energy conversion beyond conventional thermal cycles.

{In this work, we explore quantum heat engines employing qubit and qutrit working media coupled to thermal reservoirs through GAD channels. The study focuses on the role of quantum channels in describing energy exchange processes such as heat absorption, dissipation, and work generation in Quantum heat engines. Special attention is given to the effects of quantum correlations, population dynamics, and system-environment interactions on the overall thermodynamic behavior of the engine. In addition, we investigate the ergotropy of qubit and qutrit systems under dissipative evolution and analyze how environmental effects influence the amount of extractable work available in these quantum thermal machines.}

\section{Quantum heat engine  using quantum channel}
{A two-level quantum system is described by two discrete energy eigenstates: the ground state \( \lvert g \rangle \), corresponding to lower energy, and the excited state \( \lvert e \rangle \), corresponding to higher energy. The Hamiltonian governing the system is given by:}

\begin{equation}
  H =  \begin{pmatrix}
        \epsilon_g && 0   \\
        0 && \epsilon_e
    \end{pmatrix}, \label{Hamiltonian}
\end{equation}

{where \( e_g \) and \( e_e \) represent the energies of the ground and excited states, respectively. The statistical populations of these states are denoted by $p_g$ and $p_e$, respectively, where $p_g$ represents the probability of finding the system in the ground state and $p_e$ represents the probability of finding it in the excited state such that the populations satisfy the normalization condition $p_g + p_e = 1.$ The inital state is prepared as
}
\begin{equation}
  \rho =  \begin{pmatrix}
        p_g && 0   \\
        0 && p_e
    \end{pmatrix}. \label{state}
\end{equation}
{We consider a quantum heat engine whose working substance is initially prepared in the state given in Eq.~(\ref{state}). In the first stage of the cycle, the system interacts with a unitary reservoir and evolves under the action of the unitary operator $U_1$. The unitary evolution is chosen such that the populations of the energy eigenstates remain unchanged during the process. Consequently, only the coherences of the state may be modified, while the diagonal elements in the energy basis are preserved. The state of the system after this transformation can therefore be written as follows:}
\begin{equation*}
\rho_1 = U_1 \,\rho\, U_1. 
\end{equation*}
{The system is subsequently brought into contact with an external environment, allowing energy exchange that modifies the population distribution between the ground and excited states. As a consequence of this interaction, the occupation probabilities of the two energy levels are redistributed according to the properties of the surrounding reservoir. This stage of the evolution is modeled using the Generalized Amplitude Damping (GAD) channel, which effectively captures the influence of a finite-energy environment on a two-level system. The corresponding evolution is described by a set of Kraus operators, given by \citep{corltn}.}

\begin{align}
A_0 &= \sqrt{f}
\begin{pmatrix}
1 & 0\\
0 & \sqrt{1-\gamma}
\end{pmatrix},
\qquad
A_1 = \sqrt{f}
\begin{pmatrix}
0 & \sqrt{\gamma}\\
0 & 0
\end{pmatrix},
\\[6pt]
A_2 &= \sqrt{1-f}
\begin{pmatrix}
\sqrt{1-\gamma} & 0\\
0 & 1
\end{pmatrix},
\qquad
A_3 = \sqrt{1-f}
\begin{pmatrix}
0 & 0\\
\sqrt{\gamma} & 0
\end{pmatrix}\label{GAD}.
\end{align}
{Here, the parameter $\gamma = 1- e^{-\Gamma t}$
, where $\Gamma$ denotes the spontaneous transition rate and $t$ represents the interaction time. The interaction is chosen such that the population of the excited state increases as a result of energy gained from the environment. The parameter $f$
characterizes the relative likelihood of downward transitions, while $1-f$
 corresponds to upward transitions induced by the reservoir. The value of $f$ depends on the relative energetic configuration of the system and the environment, determining the direction and magnitude of population redistribution.}

{The evolved state under the GAD channel is given by the completely positive trace-preserving map:}
\begin{equation}
\rho_2 = \sum_k A_k \rho_2 A^{\dagger}_k,
\label{rep}
\end{equation}
{where $\rho_1$ is the state before this interaction. After applying the Kraus operators, the state becomes}
\begin{equation}
\rho_2 = 
\begin{pmatrix}
f \gamma p_e + [1+(f-1)\gamma] p_g && 0\\
0 && (1-f \gamma)p_e - (f-1)\gamma p_g
\end{pmatrix}.\label{evolved}
\end{equation}
For the excited-state population to exceed that of the ground state after the interaction, the following condition must be satisfied:

\begin{equation}
\frac{1+2 \gamma (f-1)}{1-2 \gamma f} < \frac{p_e}{p_g}.\label{condition_1}
\end{equation}
This inequality specifies the requirement on the initial population ratio necessary to achieve a higher occupation probability in the excited state following the environmental interaction.

Moving on to Stage III, the system is subjected to another unitary transformation governed by the operator $U_2$. During this stage, the evolution remains purely unitary, ensuring that the probability distribution among the energy levels is preserved. Consequently, the populations of the ground and excited states remain unchanged, while only the coherence properties of the state may be altered. The transformed state after this operation can be expressed as:

\begin{equation}
\rho_3 = U \rho_2 U^{\dagger}.
\end{equation}

In the final stage of the cycle, the system is brought into contact with an environment characterized by lower energy. This interaction facilitates energy transfer from the system to the surroundings, leading to an increased population in the ground state relative to the preceding stage. As a result, the population distribution shifts in favor of the lower-energy level.

This process is modeled using a Generalized Amplitude Damping (GAD) channel with appropriately chosen parameters. In this stage, the transition parameter is set to $f=1$
, corresponding to purely downward transitions, while the decoherence parameter is taken as $\gamma = 0$, ensuring controlled population dynamics during the interaction. The resulting state is given by

\begin{equation}
\rho_1 = \sum_k A'_k \rho_3 A_k'^{\dagger}.
\end{equation}
\section{Work done by a two level heat engine}
{In this section, we analyze the work extraction for a two-level quantum heat engine (QHE). Consider a two-level system interacting with hot and cold reservoirs. The occupation probabilities of the excited state corresponding to the cold and hot reservoirs are given by}
\begin{equation}
p_c = \frac{1}{1 + e^{-\beta_1 \Delta_c}},
\end{equation}
\begin{equation}
p_h = \frac{1}{1 + e^{-\beta_2 \Delta_h}},
\end{equation}
{where the subscripts $c$ and $h$ denote the cold and hot reservoirs, respectively. Here, $\beta_1 = 1/(k_B T_c)$ and $\beta_2 = 1/(k_B T_h)$ are the inverse temperatures of the cold and hot reservoirs, and $\Delta_c$, $\Delta_h$ represent the corresponding energy gaps of the two-level system.}

{The work extracted per cycle of the two-level QHE can then be expressed as}
\begin{equation}
W_1 = (p_c - p_h)(\Delta_c - \Delta_h).
\end{equation}

{The maximum work is obtained in the limiting case where the cold reservoir approaches infinite temperature ($\beta_1 \to 0$) and the hot reservoir approaches zero temperature ($\beta_2 \to \infty$). In this limit, the occupation probabilities reduce to $p_c = \frac{1}{2}$ and $p_h = 1$, yielding}
\begin{equation}
W^{(1)}_{\text{max}} = \frac{1}{2}(\Delta_h - \Delta_c).
\end{equation}

{This represents the maximum work extractable from a two-level QHE under ideal conditions.}

{We now turn to a more realistic scenario, where work extraction is performed via a quantum channel, and the system is coupled to an environment at finite temperature. In this case, the heat absorbed by the system depends on the system-environment interaction and is given by}
\begin{align}\nonumber
Q_1 &= \text{tr}(\rho_2H -\rho_1 H)\\
    &= [(1-f)\gamma p_g -f\gamma p_e]\Delta_h \geq 0.
\end{align}
For absorption to occur, $f<\frac{1}{2}$ and $p_g > p_e$, the heat absorbed is positive, i.e $Q_1> 0$.

The heat rejected in de excitation via amplitude damping channel can be represented as 
\begin{align}\nonumber
Q_2 &= \text{tr}(\rho_1H' -\rho_3 H')\\
    &= [(f-1)\gamma p_g + f\gamma p_e]\Delta_c.
\end{align}
When the ground state population is significantly higher than the excited state, or when the probability of emission is very low, the resulting quantity is always negative, indicating that $Q_2 < 0$. This can be alternatively expressed as a higher amount of heat being absorbed or rejected when the ground state is initially more populated.

The work done for the protocol using quantum channel is calculated as
\begin{align}
W &= Q_1 + Q_2\\
  &= [(1-f)\gamma p_g - f\gamma p_e](\Delta_h - \Delta_c) \geq 0.
\end{align}
The work done in this system is influenced by factors such as the probability of emission, decoherence parameter, and population density. In the ideal case where the system interacts with the reservoir asymptotically or the decay rate is sufficiently large, resulting in negligible probability of emission, the maximum work done is solely determined by the population in the ground state. This can be expressed as follows:
\begin{equation}
W_{max} = p_g(\Delta_h - \Delta_c).
\end{equation}
One can achieve the maximum work output when the initial population is predominantly concentrated on the ground state and subsequently all the particles transit to the excited state in the next state, also known as population inversion.

In short we can say $W_{max} \geq W^1_{max}$.

The positive work condition can be expressed as 
\begin{equation}
\frac{1-f}{f} \geq \frac{p_e}{p_g},
\end{equation}
the equality holds for nil work.
\begin{figure}
\includegraphics[width=0.8\linewidth]{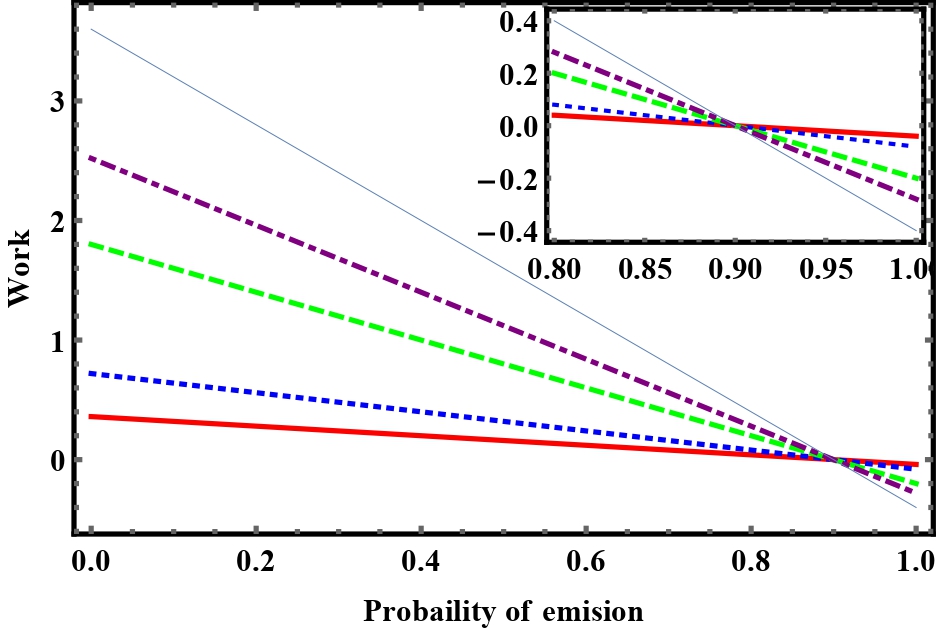}
\caption{Work extracted form the thermal machine with $\gamma =$ 0.1 (red), 0.2 (blue), 0.5 (green), 0.7 (purple), 1 (grey). Inset shows the magnified plot at $f = 0.9$.}
 \label{gad_deco_f}
\end{figure}

The probability of absorption plays a crucial role in the efficient delivery of work from a thermal machine, allowing for maximum population inversion and optimal work output when the system returns to its initial state. Figure (\ref{gad_deco_f}) illustrates the impact of the emission probability ($f$) on work extraction. It is evident that minimizing emission (maximizing absorption) leads to the highest achievable work. This implies that when the absorption probability is high, the excited state reaches its maximum population, resulting in the greatest degree of population inversion. Furthermore, a higher decoherence rate promotes energy emission from the excited state, causing the system to return to its initial condition and maximizing the performed work. It is noteworthy that positive work can only be obtained when the emission probability is less than $90\%$, beyond which work extraction becomes unfeasible.
\begin{figure}
\includegraphics[width=0.8\linewidth]{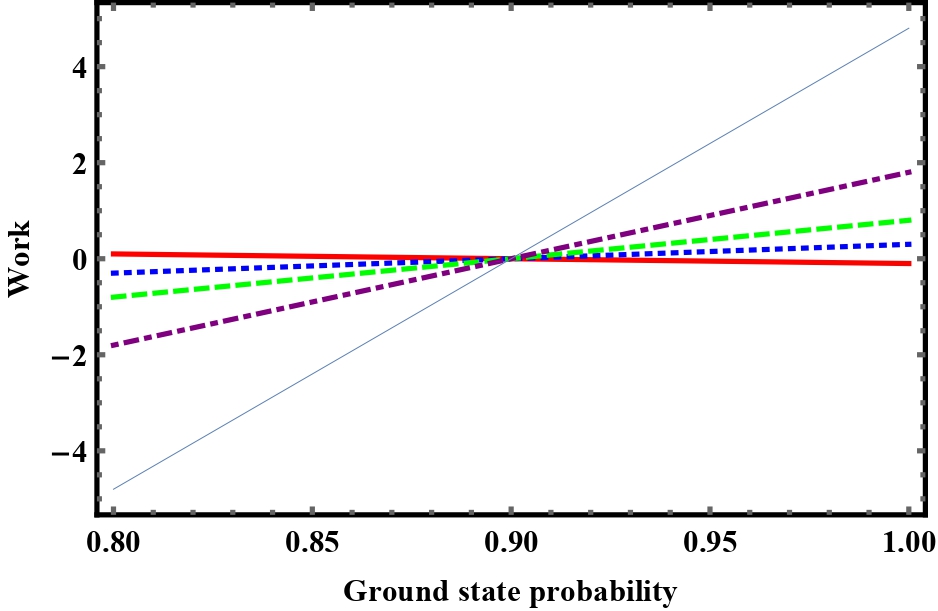}
\caption{Work extracted as a function probabilitry distribution in ground state with difference in hot bath energy $\Delta h =$ 1 (red), 5 (blue), 10 (green), 20 (purple), 50 (grey).}
\label{gad_deco_pg}
\end{figure}

In Figure (\ref{gad_deco_pg}), we illustrate the significance of the ground state probability distribution and its influence on work extraction. It is evident that positive work can only be extracted when the population of the ground state exceeds that of the excited state. Intuitively, a larger energy difference ($\Delta$) between the hot bath and the system leads to increased work output from the thermal machine. Regardless of the hot bath temperature, there must be a minimum distribution of particles in the ground state to achieve extractable positive work.

So far, we have observed that the heat engine returns to the initial state once the cycle is complete, where the system is allowed to interact with the cold bath asymptotically. However, in reality, this is an ideal case and at finite time, the system may not reach the initial state. To address this, after stage III, instead of allowing the system to reach the initial state at infinite time, we introduce an amplitude damping channel (AD) that models spontaneous emission at finite time. The Kraus operators are given by eq.(\ref{GAD}) with a probability of emission $f=1$. Due to the action of the AD channel, the state evolves to a new state different from the initial state given in eq.(\ref{state}). The evolved state is expressed as: 
\begin{equation}
\rho'
=\begin{pmatrix}
p'_g && 0 \\
0 && p'_e
\end{pmatrix}
\end{equation}
where 
\begin{equation*}
p'_g = (k + f \gamma - k f \gamma) p_e + (1+(f-1 +k -f k)\gamma)p_g
\end{equation*}
and
\begin{equation*}
p'_e= (1-k)[(1-f\gamma)p_e + (f-1)\gamma p_g]
\end{equation*}
with $k$ is the decoherance parameter in AD channel. In this scenario, the parameters can be adjusted to ensure that the ground state population ($p'_g$) is higher than that of the excited state ($p'_e$), that is, $p'_g > p'_e.$ 

The work output in quantum heat engines is primarily determined by the energy gap between the system's energy levels and the initial probability distribution in the ground state. A higher population in the ground state leads to a greater amount of extractable work. This behavior can be understood as the quantum channel facilitates population inversion by coupling the system with its environment.

\begin{figure}
\includegraphics[width=0.8\linewidth]{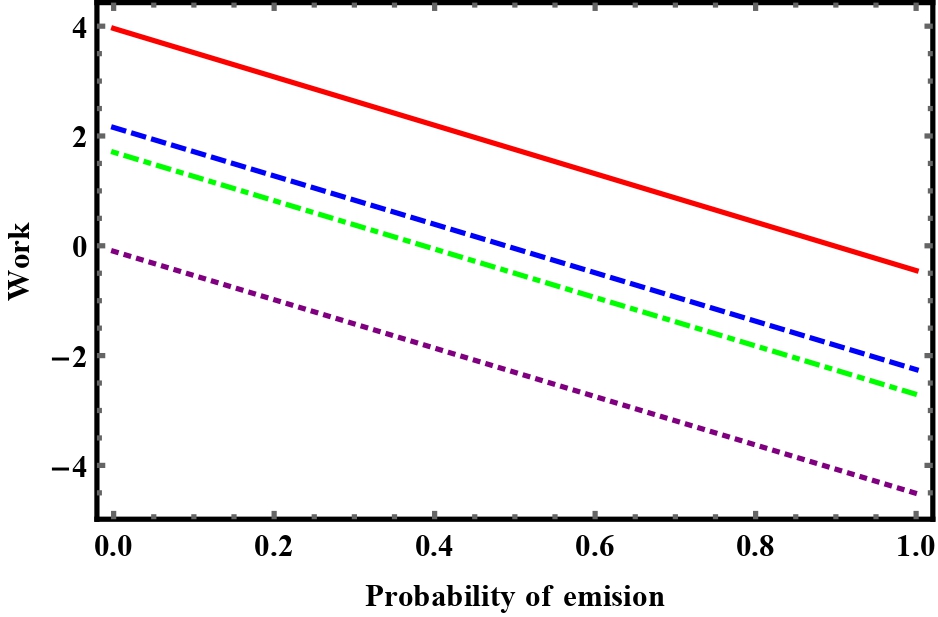}
\caption{Work as function of probability of emission with $p_g$ = 0 (Dotted), 0.4 (Dot-Dashed), 0.5 (Dashed), 0.9 (Solid).}
\label{p_g}
\end{figure}

The impact of the ground state population ($p_g$) on the extracted work is depicted in fig.(\ref{p_g}). As evident from the graph, the work decreases as the ground state population decreases and eventually reaches zero. It is worth noting that there exists a critical value of ground state population beyond which no work can be extracted. However, a small amount of work can still be obtained even when $p_g < p_e$, but this is only possible when the temperature of the environment is relatively high, corresponding to a smaller value of $f'$. In practical terms, it becomes impossible to extract any work when the population is solely concentrated in the excited state.

Here the state $\rho'$ is deviated from the initial state in eq.(\ref{state}) and the distance between these states are calculated using Hilbert-Schmidt norm given by
\begin{align*}
D(\rho,\rho')&= \lVert \rho - \rho'\rVert\\
             &= \sqrt{2}[((1-f)(1-k)\gamma )p_g - (f\gamma + k(1-f \gamma))p_e].
\end{align*}
This indicates the deviation of the state under the AD channel from the initial state. When $D(\rho, \rho') = 0$, the evolved state has returned to the initial condition, and the heat engine repeats the same cycle. On the other hand, for a non-zero $D(\cdot,\cdot)$, the heat engine follows a different cycle. It is important to note that the number of particles in the heat engine is not exchanged with the environment, resulting in the total number of particles remaining constant throughout the process.
\begin{figure}
\includegraphics[width=0.8\linewidth]{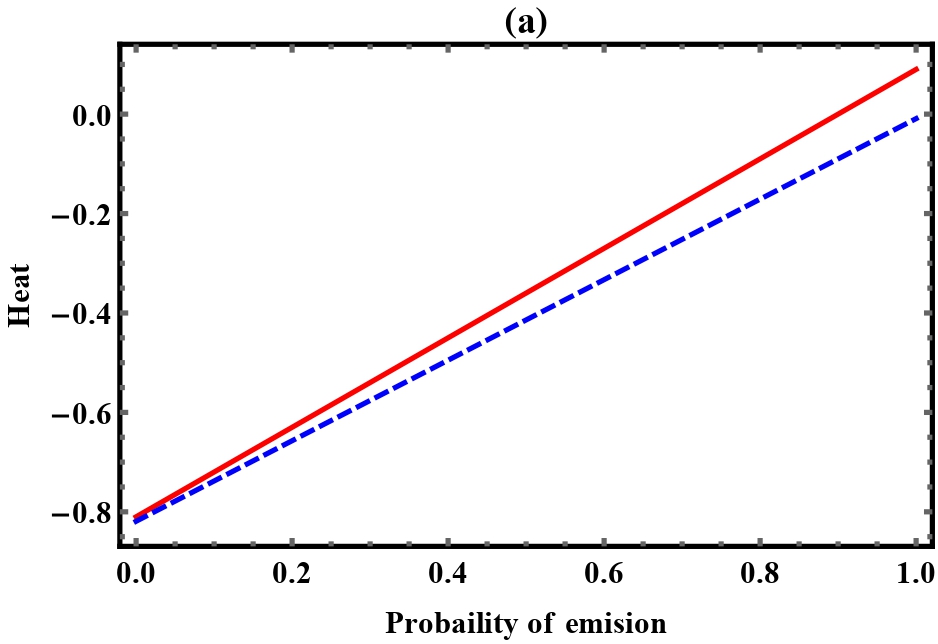}
\includegraphics[width=0.8\linewidth]{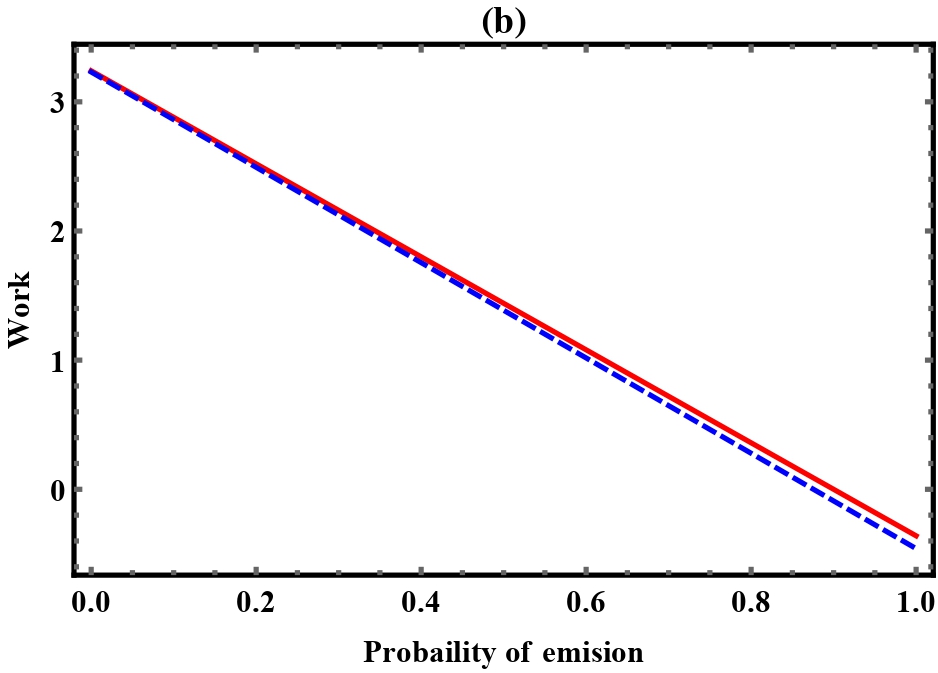}
\caption{(a) Heat rejected for cyclic (solid), non-cyclic (dashed) as a function of probaility of emission. (b) Work done for cyclic (solid), non-cyclic (dashed).}
\label{gad_deco_k}
\end{figure}

In Fig. (\ref{gad_deco_k}), we have compared the heat rejected and work done for both cyclic and non-cyclic processes. The results demonstrate that there are no qualitative differences between the two processes, with the only distinction being a slight reduction in the work done by the engine when the initial state is not reached. The additional work is utilized for the redistribution of particles among the energy levels, as represented by the following expression:
\begin{equation}
\delta W = [f \gamma + k (1-f \gamma)p_e - (1-f)(1-k)\gamma p_g]\Delta_h.
\end{equation}



\section{Qutrit quantum heat engines}
{We now extend the analysis to a qutrit system, which consists of three discrete energy levels. In thermal equilibrium, the population is distributed among these three levels according to the corresponding occupation probabilities. The density matrix of the qutrit system in the energy eigenbasis can be written as}
\begin{equation}
\tau=
\begin{pmatrix}
p_0 & 0 & 0\\
0 & p_1 & 0\\
0 & 0 & p_2
\end{pmatrix},
\label{qutrit_}
\end{equation}
{where $p_0$, $p_1$, and $p_2$ represent the occupation probabilities of the energy eigenstates associated with the energy levels $E_1$, $E_2$, and $E_3$, respectively. These probabilities satisfy the normalization condition}
\begin{equation}
p_0 + p_1 + p_2 = 1.
\end{equation}

{Similar to the two-level quantum heat engine, the qutrit working medium is first subjected to a unitary transformation, which coherently redistributes the population among the three energy levels and may generate quantum coherence in the system. The evolved state after the unitary operation is given by}
\begin{equation}
\tau_1 = U \tau U^\dagger,
\end{equation}
{where $U$ is a $3 \times 3$ unitary operator satisfying $U^\dagger U = U U^\dagger = I$.}

{After the unitary stroke, the system interacts with a thermal environment  described by the generalized amplitude damping (GAD) channel. This channel captures both relaxation and excitation processes at finite temperature, making it suitable for modeling open-system thermalization effects. The final state of the system is obtained through the action of the Kraus operators of the qutrit GAD channel.}

{The Kraus operators corresponding to the qutrit generalized amplitude damping channel are given by}
\begin{align}
F_0 &= \sqrt{f'}
\begin{pmatrix}
1 & 0 & 0 \\
0 & \sqrt{1-\lambda_1} & 0 \\
0 & 0 & \sqrt{1-\lambda_2}
\end{pmatrix}, \\[6pt]
F_1 &= \sqrt{f'}
\begin{pmatrix}
0 & \sqrt{\lambda_1} & 0 \\
0 & 0 & 0 \\
0 & 0 & 0
\end{pmatrix}, \qquad
F_2 = \sqrt{f'}
\begin{pmatrix}
0 & 0 & \sqrt{\lambda_2} \\
0 & 0 & 0 \\
0 & 0 & 0
\end{pmatrix}, \\[6pt]
F_3 &= \sqrt{f'}
\begin{pmatrix}
\sqrt{(1-\lambda_1)-\lambda_2} & 0 & 0 \\
0 & 1 & 0 \\
0 & 0 & 1
\end{pmatrix}, \\[6pt]
F_4 &= \sqrt{1-f'}
\begin{pmatrix}
0 & 0 & 0 \\
\sqrt{\lambda_1} & 0 & 0 \\
0 & 0 & 0
\end{pmatrix}, \\[6pt]
F_5 &= \sqrt{1-f'}
\begin{pmatrix}
0 & 0 & 0 \\
0 & 0 & 0 \\
\sqrt{\lambda_2} & 0 & 0
\end{pmatrix}.
\end{align}
where $\lambda_1,\lambda_2 \in [0,1]$ represents  the probabilities of $\lvert 1 \rangle \rightarrow \lvert 0 \rangle$, $\lvert 2\rangle \rightarrow \lvert 0 \rangle$ respectively. We have $\lambda_i = 1-e^{-\Gamma_i t}$ with $\Gamma$ being the spontaneous decay rate with $t$ being the time period. The parameter $f'$ is related to the temperature of the GAD channel which is given by $f' = (1+e^{-\beta E})^{-1}$ where smaller value of $f'$ corresponds to higher temperature.

{After preparing the initial state as given in Eq.~(\ref{qutrit_}), the system is allowed to interact with the environment through the generalized amplitude damping (GAD) channel. This interaction models the exchange of energy between the system and its surrounding thermal environment. The evolution of the density matrix under the GAD channel is expressed as
\begin{equation}
\tau_2 = \sum_k F_k \tau_1 F_k^\dagger,
\end{equation}
where $F_k$ represent the Kraus operators corresponding to the GAD process.}

{As a consequence of this interaction, the system absorbs energy from the environment, leading to a change in its internal energy. The heat absorbed during this stage is therefore calculated as
\begin{align*}
Q_1 &= \text{tr}[(\tau_2 - \tau_1)H_1] \\
&= (1-f')p_0[\Delta^h_{10}\lambda_1 + \Delta^h_{20}\lambda_2] \\
&\quad + f[\Delta^h_{10}\lambda_1 p_1 + \Delta^h_{20}\lambda_2 p_2].
\end{align*}
Here, $\Delta^h_{ij}$ denotes the energy gap associated with the hot reservoir, while $\lambda_1$ and $\lambda_2$ characterize the transition probabilities between different energy levels. The above expression clearly shows that the absorbed heat depends on both the energy-level spacing and the occupation probabilities of the corresponding states.}

{In the second stage of the cycle, the system undergoes a unitary evolution, which corresponds to a work extraction stroke. Since unitary evolution preserves entropy, no heat exchange takes place during this process. The evolved state after the unitary operation is given by
\begin{equation}
\tau_3 = U \tau_2 U^\dagger,
\end{equation}
where $U$ is the unitary operator governing the coherent evolution of the system.}

{Following the unitary stroke, the system is again coupled to the environment through the GAD channel. In this stage, the probability of emission is taken as $f' = 1$, allowing the system to dissipate energy into the surroundings. The final state after this interaction is expressed as
\begin{equation}
\tau_4 = \sum_k F_k \tau_3 F_k^\dagger.
\end{equation}}

{The heat rejected by the system during this process is calculated as
\begin{align*}
Q_2 &= \text{tr}[(\tau_1 - \tau_3)H_2] \\
&= (1-f')p_0[\Delta^c_{01}\lambda_1 k_1 + \Delta^c_{12}\lambda_2 k_2] \\
&\quad + k_1p_1(1-f\lambda_1)\Delta^c_{01} \\
&\quad + k_2p_2(1-f\lambda_2)\Delta^c_{02}.
\end{align*}
Here, $\Delta^c_{ij}$ represents the energy gaps associated with the cold reservoir, while $k_1$ and $k_2$ denote the contributions arising from transitions involving the excited states. This expression indicates that the rejected heat is determined by the spontaneous emission processes occurring between different energy levels of the qutrit system.}

{In general, $\Delta_{ij}$ denotes the energy difference between the $i$th and $j$th energy levels. The total work extracted during the complete cycle is obtained from the net heat exchange and is given by
\begin{equation*}
W = Q_1 + Q_2.
\end{equation*}}

\begin{figure}
\includegraphics[width=0.8\linewidth]{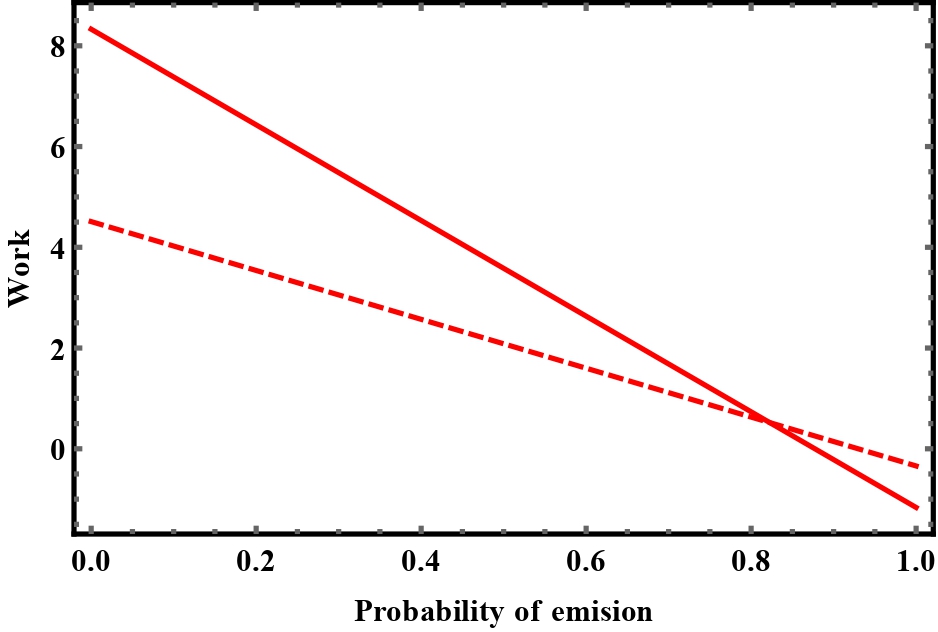}
\caption{Work done as a function of the probability of emission for qubit (dotted) and qutrit (solid) systems.}
\label{gad_qutrit}
\end{figure}

{The work output strongly depends on the energy-level structure of the system and the population distribution among the different states. In particular, the ground-state population and the transition probabilities play a crucial role in determining the amount of extractable work. By carefully tuning the energy gaps and controlling the interaction parameters, the work output can be significantly optimized.}

{Unlike qubit systems, qutrit systems possess multiple excited states, which provide additional transition channels for energy exchange. These extra pathways enhance the contribution from spontaneous emission processes and consequently improve the overall work extraction capability. This enhancement is clearly observed in Fig.~(\ref{gad_qutrit}), where the qutrit system exhibits higher work output compared to the qubit system for the same energy gap configuration.}

{The improved performance of the qutrit system arises primarily due to the participation of higher excited states in the thermodynamic cycle. The spontaneous emission from both the first and second excited states contributes constructively to the extracted work, leading to a noticeable increase in efficiency and work output. However, it is important to note that the final state of the qutrit system after the completion of the cycle is generally different from its initial state, indicating that the process does not necessarily correspond to a perfectly cyclic evolution.}

\section{Efficiency}
{In quantum heat engines, the efficiency is primarily governed by the energy gap between the hot and cold reservoirs, analogous to the dependence of Carnot efficiency on the reservoir temperatures. In such systems, the working medium returns to its initial state at the end of each cycle, and therefore the efficiency is predominantly determined by thermodynamic parameters associated with the energy-level structure of the system.}

{In contrast, the efficiency of non-cyclic heat engines depends on several additional factors beyond the energy gaps. These include the initial probability distribution of the quantum states, the decoherence parameters associated with the environmental interaction, and the probability of emission governing the dissipative dynamics. Since the system does not necessarily return to its initial state after the completion of the process, the efficiency becomes strongly influenced by the dynamical evolution of the state populations during the interaction with the environment.}

{Among these parameters, the initial ground-state probability distribution plays a particularly significant role. The population distribution determines how energy is absorbed and emitted during the thermodynamic cycle and therefore directly influences the amount of extractable work. A higher ground-state population generally modifies the transition probabilities between different energy levels, thereby affecting both the heat exchange processes and the overall engine performance. Consequently, even small variations in the initial probability distribution can lead to substantial changes in the work output and efficiency.}

{Although both qubit and qutrit systems exhibit qualitatively similar efficiency behavior, their quantitative performance differs considerably. The qutrit system benefits from the presence of additional excited states, which enhance the work extraction capability through multiple transition pathways. However, despite the increased work output, the qubit system is observed to possess comparatively higher efficiency under certain parameter regimes. This difference arises because the additional transitions in the qutrit system also introduce extra dissipation channels, which can reduce the ratio of useful work to absorbed heat.}

{Furthermore, there exists a critical region in the probability of emission where the efficiency decreases abruptly. This sudden reduction in efficiency is closely related to the redistribution of state populations caused by the environmental interaction. The location and extent of this transition region strongly depend on the initial probability distribution of the system. As the emission probability increases, the balance between heat absorption and heat dissipation changes significantly, leading to a rapid decline in the efficiency beyond a certain threshold.}

\begin{figure}
\includegraphics[width=0.8\linewidth]{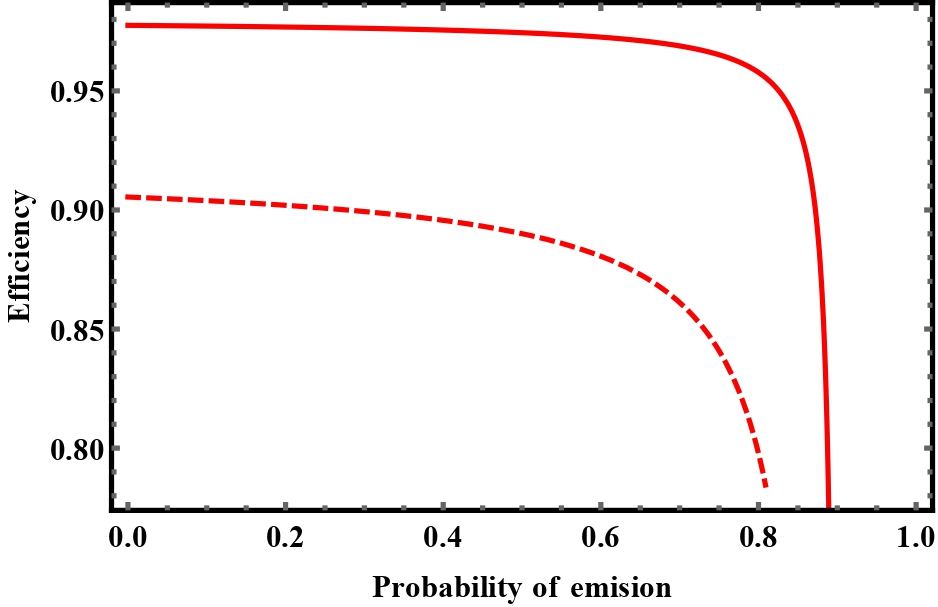}
\caption{Efficiency for qubit (Solid) and qutrit (Dotted) system.}
\label{efficiency}
\end{figure}

{This behavior is illustrated in Fig.~(\ref{efficiency}), where the dependence of efficiency on the probability of emission is presented for different initial probability distributions. The figure clearly demonstrates the sensitivity of the engine performance to both the emission probability and the initial state configuration.}

\section{Ergotropy Evaluation via Kraus Dynamics}
We now investigate the ergotropy of two-level (qubit) and three-level (qutrit) quantum heat engines operating under dissipative thermal environments. Ergotropy represents the maximum amount of work that can be extracted from a quantum system through cyclic unitary operations while keeping the system Hamiltonian unchanged~\cite{allahverdyan2004maximal}. Unlike classical extractable work, ergotropy captures the role of quantum population distribution and coherence in determining the usable energy stored within a quantum state.

\subsection{Qubit state}
We consider a two-level working medium described by the Hamiltonian
\begin{equation}
H_S=\tfrac{\hbar\omega}{2}\sigma_z,
\end{equation}
with energy eigenstates $\{|g\rangle,|e\rangle\}$ and eigenvalues
$E_g=-\tfrac{\hbar\omega}{2}$ and $E_e=\tfrac{\hbar\omega}{2}$.

The interaction with a thermal environment is modeled through a generalized amplitude damping (GAD) channel. The system evolution is described by the Kraus map Eq.\ref{rep}
where the Kraus operators satisfy the completeness relation
\begin{equation}
\sum_{i=0}^{3}A_k^\dagger A_k=\mathds{I}.
\end{equation}

The channel is characterized by a temperature parameter $f$ and a time-dependent damping strength
\begin{equation}
\lambda(t)=1-e^{-\gamma t},
\end{equation}
which quantifies the strength of system–bath interaction.

For an initially diagonal state
\begin{equation}
\rho(0)=\mathrm{diag}(p_g(0),p_e(0)),
\end{equation}
the density matrix remains diagonal during evolution,
\begin{equation}
\rho(t)=\mathrm{diag}(p_g(t),p_e(t)).
\end{equation}

The excited-state population evolves as
\begin{equation}
p_e(t)=p_e(0)\big(1-f\lambda(t)\big)
       +p_g(0)\big(1-f\big)\lambda(t).
\end{equation}

The instantaneous energy of the working medium is
\begin{equation}
E(t)=\mathrm{Tr}\!\big[\rho(t)H_S\big]
     =\tfrac{\hbar\omega}{2}\big(p_e(t)-p_g(t)\big).
\end{equation}

Ergotropy is defined as the maximum amount of work that can be extracted from a quantum state through cyclic unitary operations, i.e., operations that leave the system Hamiltonian unchanged~\cite{allahverdyan2004maximal,campaioli2019quantum}. It is given by
\begin{equation}
\mathcal{W}(t)=
\mathrm{Tr}\!\big[\rho(t)H_S\big]
-
\mathrm{Tr}\!\big[\rho_{\mathrm{pas}}(t)H_S\big],
\end{equation}
where $\rho_{\mathrm{pas}}(t)$ denotes the passive state associated with $\rho(t)$. The passive state is constructed by rearranging the eigenvalues of $\rho(t)$ in non-increasing order and assigning them to ascending energy levels. Physically, this corresponds to placing the largest population in the lowest-energy state, thereby minimizing the system energy under unitary transformations \citep{Binder2015}.

For a qubit this yields
\begin{equation}
\mathcal{W}(t)=
\begin{cases}
0, & p_g(t)\ge p_e(t),\\[6pt]
\hbar\omega\,[p_e(t)-p_g(t)], & p_e(t)>p_g(t).
\end{cases}
\end{equation}

Thus, ergotropy is nonzero only when population inversion occurs, i.e., when the excited-state population exceeds that of the ground state. In the channel framework, this inversion is governed by the balance between emission and absorption processes encoded in the temperature parameter $f$.
\begin{figure}
    \centering
    \includegraphics[width=0.8\linewidth]{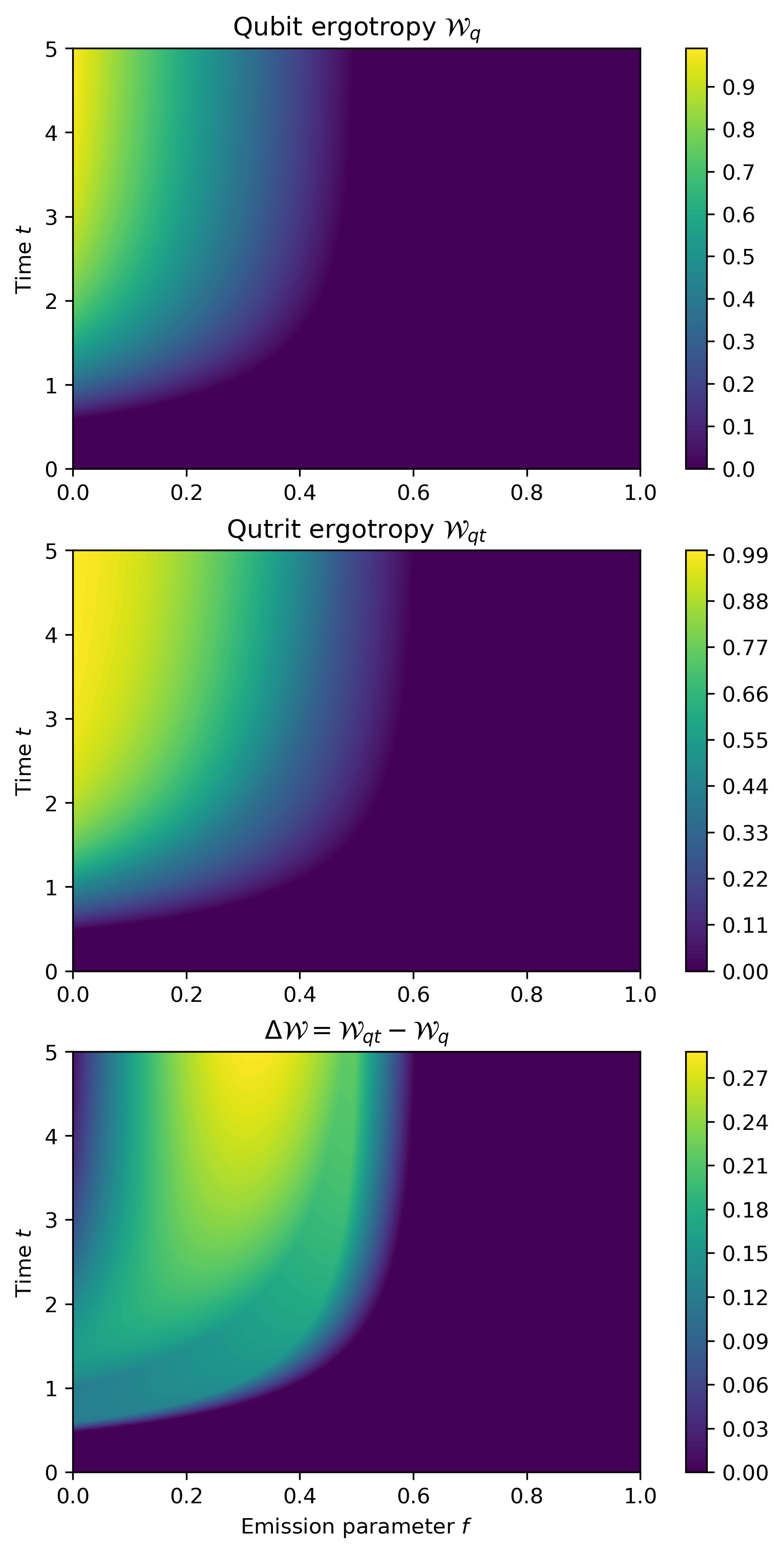}
    \caption{Ergotropy $\mathcal{W}(f,t)$ for qubit(top) and qutrit(middle) working media under time-dependent damping $\lambda_{1,2}=1-e^{-\gamma_{1,2}t}$. The difference map $\Delta\mathcal{W}$(bottom) highlights the enhanced extractable work enabled by multilevel energy structures across a wide operational parameter space.}
    \label{fig:qt}
\end{figure} 
To visualize the operational regimes for extractable work, we compute the ergotropy landscape as a function of the temperature parameter $f$ and interaction time $t$ Fig.\ref{fig:qt}(top). Here $f$ quantifies the relative weight of emission versus absorption in the thermal channel, while the damping strength is governed by the time-dependent parameter $\lambda(t)$. The resulting heat map, shown in Fig., reveals that ergotropy increases with interaction time under absorption-dominated conditions (low $f$), reflecting progressive population transfer to the excited state. Conversely, for large $f'$ emission dominates, driving the system toward a passive ground-state configuration with vanishing ergotropy.

\subsection{Extension to Qutrit Systems}

The ergotropy framework naturally extends to qutrit working media. We consider an initial diagonal state
\begin{equation}
\rho(0)=\mathrm{diag}(p_0,p_1,p_2),
\qquad
p_0+p_1+p_2=1,
\end{equation}
with system Hamiltonian
\begin{equation}
H_S=\sum_{i=0}^2 \epsilon_i |i\rangle\langle i|,
\qquad
\epsilon_0<\epsilon_1<\epsilon_2.
\end{equation}

The interaction with a thermal reservoir is modeled through a three-level generalized amplitude damping (GAD) channel characterized by time-dependent decay probabilities
\begin{equation}
\lambda_1(t)=1-e^{-\gamma_1 t}, 
\qquad
\lambda_2(t)=1-e^{-\gamma_2 t},
\end{equation}
associated with decay processes $|1\rangle\!\to|0\rangle$ and $|2\rangle\!\to|0\rangle$, respectively. The temperature parameter $f$ determines the relative strength of emission versus absorption processes.

Since the channel preserves diagonal structure, the evolved state remains
\begin{equation}
\rho(t)=\mathrm{diag}\big(p_0(t),p_1(t),p_2(t)\big).
\end{equation}

The instantaneous energy of the working medium is
\begin{equation}
E(t)=\mathrm{Tr}\!\big[\rho(t)H_S\big]
     =\sum_{i=0}^2 \epsilon_i\,p_i(t).
\end{equation}

To determine the extractable work, we construct the passive state $\rho_{\rm pas}(t)$, defined as the state obtained by rearranging the eigenvalues of $\rho(t)$ in non-increasing order and assigning them to ascending energy levels. Denoting the sorted populations as
\begin{equation}
p_0^{\downarrow}(t)\ge p_1^{\downarrow}(t)\ge p_2^{\downarrow}(t),
\end{equation}
the passive state reads
\begin{equation}
\rho_{\rm pas}(t)=
\sum_{i=0}^2 p_i^{\downarrow}(t)\,|i\rangle\langle i|.
\end{equation}

Its associated energy is therefore
\begin{equation}
E_{\rm pas}(t)=
\sum_{i=0}^2 \epsilon_i\,p_i^{\downarrow}(t).
\end{equation}

Ergotropy is defined as the maximum work extractable via unitary operations, given by the difference between the actual and passive energies,
\begin{equation}
\mathcal{W}(t)=E(t)-E_{\rm pas}(t).
\end{equation}

For diagonal $\rho(t)$, both $\rho(t)$ and $\rho_{\rm pas}(t)$ commute with $H_S$, and the unitary that extracts maximal work corresponds to a permutation of populations. Thus,
\begin{align}
\mathcal{W}(t)
&=\sum_{i=0}^2 \epsilon_i\,p_i(t)
 -\sum_{i=0}^2 \epsilon_i\,p_i^{\downarrow}(t) \\
&=\sum_{i=0}^2 \epsilon_i
   \big[p_i(t)-p_i^{\downarrow}(t)\big].
\end{align}

Since $\{\epsilon_i\}$ are ordered increasingly while $\{p_i^{\downarrow}\}$ are ordered decreasingly, the rearrangement inequality ensures
\begin{equation}
E_{\rm pas}(t)\le E(t),
\end{equation}
with equality holding only when the populations are already ordered passively. Hence,
\begin{equation}
\mathcal{W}(t)\ge 0,
\end{equation}
and ergotropy vanishes if and only if $\rho(t)$ is passive.

Unlike the qubit case, ergotropy in qutrit systems does not require complete inversion of a single excited state. Any population redistribution that violates passive ordering (e.g., $p_2>p_1$ or $p_1>p_0$) produces finite extractable work. Consequently, partial occupation of higher excited levels suffices to generate nonzero ergotropy, enhancing both storage capacity and robustness against dissipation.

\medskip

The corresponding ergotropy landscape for the qutrit working medium is presented in Fig.\ref{fig:qt}(middle). In contrast to the qubit case, finite ergotropy persists over a significantly broader parameter region. This enhancement originates from the multilevel energy structure: extractable work does not require complete inversion of a single excited level. Instead, partial population of higher excited states already generates a mismatch with the passive configuration, enabling nonzero ergotropy.

To highlight this multilevel advantage, Fig\ref{fig:qt}(bottom). depicts the difference map
\begin{equation}
\Delta\mathcal{W}=\mathcal{W}_{\mathrm{qutrit}}-\mathcal{W}_{\mathrm{qubit}}.
\end{equation}
The positive regions indicate operational regimes where qutrit working media retain extractable work even after qubit systems have relaxed to passive states. This demonstrates the enhanced storage capacity and dissipation resilience of higher-dimensional quantum batteries originate from their multilevel energy structure, which enables extractable work without requiring complete population inversion as been studied earlier too~\cite{Binder2015,Campaioli2017}.

\section{Summary and Conclusions}
This work presents a comprehensive study of quantum heat engines employing qubit and qutrit working media interacting with thermal environments through GAD channels. The proposed framework utilizes quantum channels to model heat absorption, dissipation, and work extraction in open quantum thermodynamic systems, enabling a detailed analysis of the role of environmental interactions, population redistribution, and dissipative dynamics in quantum thermal machines. For qubit heat engines, the study derives the conditions required for positive work extraction and demonstrates that the extracted work strongly depends on the interplay between emission probability, decoherence strength, and the initial population distribution. The analysis further shows that population inversion plays a central role in enabling useful work extraction, while finite-time dissipative effects modify the thermodynamic cycle through redistribution of state populations.

The work is further extended to qutrit heat engines, where the presence of multiple excited states introduces additional pathways for energy exchange and significantly enhances the work extraction capability of the system. The multilevel structure of qutrit systems allows greater robustness against decoherence and enables the retention of finite extractable work over a broader range of environmental parameters compared to two-level systems. In addition, the ergotropy of both qubit and qutrit systems is investigated under dissipative evolution. The results reveal that, while qubit systems exhibit nonzero ergotropy only under complete population inversion, qutrit systems can sustain finite ergotropy even under partial population redistribution due to the richer multilevel energy structure. Consequently, qutrit working media demonstrate improved energy-storage capability and enhanced resilience against environmental dissipation.

Overall, the results establish generalized amplitude damping channels as an effective framework for analyzing dissipative quantum heat engines and highlight the crucial role of quantum population dynamics, decoherence, and multilevel energy structures in determining the thermodynamic performance of quantum thermal machines. The findings provide deeper insight into the interplay between open-system dynamics and extractable work and may contribute toward the development of realistic quantum thermal devices, quantum batteries, and advanced energy-conversion technologies operating in the quantum regime.


\end{document}